\documentclass{JHEP3}
\usepackage{amsmath}
\usepackage{graphicx}
\usepackage{amsfonts}
\usepackage{amssymb}
\newcommand{\bea}{\begin{eqnarray}}
\newcommand{\eea}{\end{eqnarray}}
\newcommand{\beq}{\begin{equation}}
\newcommand{\eeq}{\end{equation}}
\newcommand{\nn}{\nonumber}

\newcommand{\lishi}{\langle\!\langle}
\newcommand{\rishi}{\rangle\!\rangle}

\newcommand{\abar}{{\overline a}}
\newcommand{\alphabar}{{\overline \alpha}}
\newcommand{\Lbar}{{\overline L}}
\newcommand{\Wbar}{{\overline W}}

\newcommand{\zbar}{{\overline z}}
\newcommand{\phibar}{{\overline \phi}}

\newcommand{\C}[1]{{\mathcal{#1}}}

\newcommand{\R}[1]{{\mathrm{#1}}}

\newcommand{\half}{{\frac{1}{2}}}
\newcommand{\quarter}{{\frac{1}{4}}}

\newcommand{\cbytwelve}{{\frac{c}{12}}}
\newcommand{\cbytwfo}{{\frac{c}{24}}}

\newcommand{\Ham}{\half(L_0+\Lbar_0-\cbytwelve)}
\newcommand{\ket}[1]{\,\vert #1\rangle}
\newcommand{\bra}[1]{\langle #1\vert\,}
\newcommand{\Tr}{\R{Tr}}
\newcommand{\ishiket}[1]{\vert\, #1\,\rishi}
\newcommand{\ishibra}[1]{\lishi\, #1\,\vert}

\preprint{OUTP-04/05P}

\title{Boundary states and broken bulk symmetries in $WA_r$ minimal models }
\author{Alexandre F. Caldeira\\Department of Physics, University of Oxford \\
Theoretical Physics,\\
1 Keble Road,\\
 Oxford OX1 3NP, UK\\
E-mail: \email{caldeira@thphys.ox.ac.uk}}

\author{John F. Wheater\thanks{Currently on leave at Niels Bohr Institute,
Blegdamsvej 17,  Copenhagen DK-2100, Denmark}\\ Department of Physics,
University of Oxford \\
Theoretical Physics,\\
1 Keble Road,\\
 Oxford OX1 3NP, UK\\
E-mail: \email{j.wheater@physics.ox.ac.uk}}
\abstract{
We study the boundary states of  $(p',p)$ rational conformal field theories
 having a $W$ symmetry
of the type $A_r$ using the multi-component free-field formalism.
 The classification of primary fields for these models
 given in the literature  is shown to be incomplete; 
we give the correct classification by demanding modular covariance and show
 that the resulting modular $S$ matrix satisfies all the necessary conditions.
 Basis  states satisfying the boundary conditions are found in the
 form of coherent
 states and as expected we find that $W$ violating states can be found for 
all these models.  We 
 construct consistent physical 
boundary for all the  rank 2
 $(p+1,p)$ models (of which the
already known case of the 3-state Potts model is the simplest example) and
find  that
 the $W$ violating  sector possesses a direct analogue of the Verlinde formula.

}

\keywords{Boundary Quantum Field Theory, Conformal and W symmetry}

\begin{document}

\section{Introduction}

The  Coulomb gas formalism \cite{Dotsenko:1984nm, Dotsenko:1985ad} provides a
powerful 
method for calculating correlation functions and conformal blocks in minimal
rational  
conformal field theories (CFTs) and boundary CFTs have been of great interest 
since Cardy's famous paper \cite{Cardy:1984bb}.
Recently it has been shown that free-field  representations may be extended from 
bulk CFTs to 
systems with boundary(ies) \cite{Kawai:2002vd,Kawai:2002pz} in the case of the
Virasoro diagonal 
minimal models and for the simplest non-diagonal case, the three state Potts
model, where a multi-component 
Coulomb Gas formalism is required
\cite{Caldeira:2003zz}. The boundary states appear as coherent states in the
free-field formalism.

The three state Potts model is of particular interest because the conformal
field theory describing
its critical point is the simplest in which there is a higher dimensional chiral
operator $W^{(3)}$ of 
dimension 3 \cite{Fateev:1987vh} . There are six boundary states originally
found 
by Cardy in which the $W^{(3)}$ current is conserved  at the boundary  
\cite{Cardy:1989ir}
but in addition  there are known to be  two more states in which the $W$ current
is not conserved.
 Affleck et al \cite{Affleck:1998nq} used fusion methods to establish these
states
 while recently in \cite{Caldeira:2003zz} it was shown that precisely these
states, and no others, 
appear also in the free field formulation. This is all consistent with the
general arguments given in
\cite{Fuchs:1998qn} that there should be precisely eight conformally invariant
boundary states in this model.

The free field formulation for the Potts model is just the simplest case of a
whole family
of $W$ minimal models whose $r$ component free field representations  are built
on the Lie algebra
 $A_r$ and are further characterised
by two relatively prime integers $p'$ and $p$. These models, denoted
$W_{r+1}(p',p)$, 
have higher dimensional chiral operators $W^{(K)}$, $K=3,\ldots r+1$ and thus an
extended symmetry algebra
of which the Virasoro algebra is a sub-algebra. The Potts model corresponds to
$W_{3}(5,4)$ and is the only 
member of the family which is also a Virasoro minimal model. It is to be 
expected that all the models will
have  boundary states which violate the higher symmetry as well as those that
conserve it. The aim of this paper 
is to extend \cite{Caldeira:2003zz} and to study these boundary states.
 For general reviews of CFTs with $W$ algebras the reader should consult
\cite{Bilal:1991eu,Bouwknegt:1993wg,Ketov:1995yd}.

This paper is organized as follows.
We start in section \ref{PRELIM} with a collection of definitions and some
results from
  the standard Coulomb Gas formalism that we need.
In section \ref{WMIN} we describe the classification of fields and the Felder
complex for $W$ minimal models
and in section \ref{COHSTATES} explain how to construct
coherent state
 representations satisfying the Virasoro boundary conditions. Section
\ref{COUPLED} deals with the 
calculation of cylinder amplitudes and establishing which of the coherent states
found previously
are coupled to the bulk physics. Modular covariance of the cylinder amplitudes
is used
in section \ref{MODPROPS} to find the classification of the primary fields; this
is 
different from that given   in the literature which appears to be incomplete. In
 section \ref{PHYSICAL} we consider the physical boundary states and annulus
partition functions.
The conservation or otherwise of $W$ currents by the different possible boundary
states is considered in
section \ref{WCURRENTS} and,  
finally, in section \ref{DISCUSS} we discuss some open issues.

\section{Preliminaries\label{PRELIM}}

The usual Coulomb gas formalism \cite{Dotsenko:1984nm, Dotsenko:1985ad}
 can be extended to CFTs with a
larger symmetry than the Virasoro algebra by introducing a multiple component
scalar field \cite{Fateev:1987vh} 
 $\Phi^j(z,\bar z)$, $j=1\ldots r$, which is a 
vector in the root space of a finite dimensional Lie
 Algebra $\mathcal A$ of rank
$r$. In this paper we will be mainly concerned with 
 the algebra $A_r$ and so will specialize to it straight away.
 Let us first fix some notation. The simple roots will be denoted
$e_j, j=1\ldots r$, and the corresponding dual weights
 $\omega_j, j=1\ldots r$.
We will use ``$\cdot$'' to denote multiplication of 
vectors and matrices in the 
root space. So the scalar product of two vectors
$u$ and $v$ in the root space will be written $u\cdot v$,
 the product of two matrices
$m_1\cdot m_2$ and so on. The simple roots and
dual weights then satisfy
\begin{equation}
e_j \cdot e_j=2,\quad  e_j \cdot e_{j+1}=-1,
 \quad e_j \cdot\omega_i =\delta_{i,j}.
\end{equation}
The positive roots are given by
\beq e_{jk}=e_j+\ldots+e_k,\quad 1\le j< k \le r.\eeq
The Weyl vector $\rho$ is defined as
\begin{equation}
\rho= \sum_{j=1}^r \omega_j ,
\end{equation}
its square is
\begin{equation}
\rho ^2 = \frac{1}{12}r(r+1)(r+2),
\end{equation}
and the fundamental weights $h_K,$ $K=1\ldots r+1,$ satisfy
\begin{eqnarray}
h_1&=&\omega_1,\nonumber\\
h_K-h_{K+1}&=&e_K.
\end{eqnarray}
We denote the  Weyl group of  $\mathcal A$ by  $\mathcal W$, an element of
it by  $w$, and let $\varepsilon_w=\det w$.
The longest element of the Weyl group, $w_0$
 is the unique element of $\C W$ that 
maps the positive roots onto the negative roots. On the simple
 roots, dual weights and fundamental weights $w_0$ has the action
\bea w_0\, e_i&=&-e_{r-i+1},\nn\\
 w_0 \,\omega_i&=&-\omega_{r-i+1}\nn\\
 w_0 \,h_K&=&h_{N+1-K},\label{Sh}\eea
and  we define its matrix representation in the root basis, $S$, by
\beq S_{ij}=-\delta_{i+j,r+1}.\eeq
Finally $I$ denotes the identity matrix.

The action for $\Phi$ takes the usual form
\begin{equation}
\mathcal{S}[\Phi]=\frac{1}{8\pi}\int d^{2}z\sqrt{g}\,\left(\partial_{\mu}%
\Phi\cdot \partial^{\mu}\Phi+4 i\alpha_{0}\rho\cdot\Phi R\right), \label{s}%
\end{equation}
where $R$ is the scalar curvature, $g$ the metric, and $\alpha_0$ a constant.
We now split $\Phi$ into a holomorphic component $\phi(z)$ and an 
anti-holomorphic component $\bar\phi(\bar z)$.
The field $\phi$ has mode expansion
\begin{equation}
\phi^{j}(z)=\phi_{0}^{j}-ia_{0}^{j}\ln z+i\sum_{n\neq0}\frac{a_{n}^{j}%
}{n}z^{-n}, \label{eq. phimode}%
\end{equation}
and similarly for  $\bar\phi$.
Canonical quantization gives the usual commutation relations
\begin{align}
\lbrack a_{m}^{j},a_{n}^{l}]  &  =m\delta^{jl}\delta_{m+n,0}%
,\nn\\
\lbrack\phi_{0}^{j},a_{0}^{l}]  &  =i\delta^{jl}. \label{heisenberg}%
\end{align}
Variation of the action with respect to the metric yields the
energy-momentum tensor
\begin{equation}
T(z)=-2\pi T_{zz}=-\frac{1}{2}:\partial\phi\cdot\partial\phi
:+2\,i\,\alpha_{0}\,\rho\cdot\partial^{2}\phi, \label{T}%
\end{equation}
which has the usual expansion 
\begin{equation}T(z)=\sum_{n\in{\mathbb Z}}%
L_{n}z^{-n-2},\end{equation}
where the operators
\begin{align}
L_{n}  &  =\frac{1}{2}\sum_{m\in{\mathbb Z}}  :a_{m}\cdot a_{n-m}:
  -2\alpha_{0}(n+1)\rho\cdot a_{n}
\end{align}
obey the Virasoro algebra with
 central charge
\begin{equation}
c=r-48\ \alpha_{0}^{2}\rho^2.
\end{equation}

Fock spaces $\C F_\alpha$ are  labeled by a vacuum $|\alpha\rangle$, 
which is an
eigenvector of the $a_{0}^{j}$ operator, and annihilated by the positive
modes
\begin{eqnarray}a_{0}^{j}|\alpha\rangle&=&\alpha^{j}|\alpha\rangle,\nonumber\\
a_{n}^{j}|\alpha\rangle&=&0,\quad n>0.
\end{eqnarray}
The Fock space is formed by applying the creation
operators to the vacuum,
\begin{equation}
 a_{-n_{1}}^{j_{1}}a_{-n_{2}}^{j_{2}}...a_{-n_{p}%
}^{j_{p}}|\alpha\rangle,
\end{equation}
 and different Fock spaces are related by 
\begin{equation}
 e^{i\beta\cdot\phi_{0}}|\alpha\rangle
=|\beta+\alpha\rangle.
\end{equation}
The chiral vertex operators $V_{\alpha}(z)$ are defined by 
\begin{equation}
V_{\alpha}(z)=:e^{i\alpha\cdot\phi(z)}:,
\end{equation}
and have  conformal dimension given by 
\begin{equation}
h(\alpha
)=\frac{1}{2}\alpha\cdot(\alpha-4\alpha_{0}\rho).\label{Halpha}
\end{equation}

\section{W minimal models\label{WMIN}}

The $W_N(p',p)$ minimal models are defined for relatively prime integers
$p'$ and $p$ such that  $p'>p>N$ 
by
\begin{eqnarray} 2\alpha_{0}=\frac{p^{\prime}-p}{\sqrt{pp^\prime}},\quad
\alpha_{+}=\frac{p^\prime}{\sqrt{pp^\prime}\,},\quad
\alpha_{-}=-\frac{p}{\sqrt{pp^\prime}}
\end{eqnarray}
and have central charge
\beq c=r\left(1-\frac{(p^{\prime}-p)^2}{{pp^\prime}}(r+1)(r+2)\right).
\eeq
Considerations  requiring a consistent  fusion algebra  lead to
the allowed values of $\alpha$ \cite{Bilal:1991eu}
\bea \alpha&=&2\alpha_0\rho-\frac{1}{\sqrt{pp^\prime}}\,\lambda(m,n),\nn\\
\lambda(m,n)&=&-pm^i\omega_i+p'n^i\omega_i,\label{alphadef}\eea
where summation over repeated $i$ is implied and the $m^i $ and $n^i$ 
are positive integers satisfying
\beq \sum_im^i<p',\quad\sum_in^i<p.\label{mnrange}\eeq
%\end{document}
So $m^i\omega_i$ and $n^i\omega_i$ are dominant weights and $\lambda(m,n)$
is a non-zero weight, although not necessarily dominant.
It is not really clear what these restrictions mean. As we discuss
 directly below there is considerable degeneracy in this set-up
 and the appearance of copies related by Weyl transformation is to be
expected. For example in $W_3(5,4)$, which is the critical three
 state Potts model and should have 6 primary fields, there  are 18 $\lambda$s
satisfying these constraints; but  $\vert\C W\vert=6$  and we get only three
copies for each primary
\footnote{This ambiguity resulted in the authors of
\cite{Caldeira:2003zz} having to take an apparently arbitrary choice of
$\lambda$s on which to build the boundary states in this model.}.
 That something is amiss is even clearer if we
look at  $W_4(6,5)$ for which there are 40 $\lambda$s satisfying 
(\ref{mnrange}) yet $\vert\C W\vert=24$.
We will resolve these puzzles by considering the modular properties
 of the theory in section \ref{MODPROPS}. For the moment it is sufficient that
 the $\lambda$s certainly can be written in the form (\ref{alphadef}).

As mentioned above there is some degeneracy in the $\alpha$s. Defining
\bea \alpha^*&=&4\alpha_0\rho-\alpha,\nn\\
\alpha_w&=&2\alpha_0\rho-\frac{1}{\sqrt{pp\prime}}w\,\lambda(m,n),
\quad w\in\C W\eea
it follows that
\beq h(\alpha)=h(\alpha^*)=h(\alpha_w).\eeq
There are then two types of representation
\begin{enumerate}
\item $\alpha^*\in\{\alpha_w\}$. This implies that
\beq \lambda(m,n)=-w\, \lambda(m,n)\eeq
for some $w\in\C W$. Since $\lambda(m,n)$ is a weight we can use the property
that only for self-conjugate representations of SU($N$)is the
weight $\lambda$  in the Weyl orbit of $-\lambda$. In this case 
there is just one self-conjugate primary field of conformal weight
 $h(\alpha)$. The highest weight in a self-conjugate representation
 is given by
\beq \lambda(m,n)=\lambda^i\omega_i,\quad \lambda^i=\lambda^{N-i}.
\label{hwsconj}\eeq
 It is clear that for $\lambda(m,n)$  to be self-conjugate
either both or neither of $m^i\omega_i$ and $n^i\omega_i$ must be 
 so. In fact the later case is excluded;
(\ref{hwsconj}) leads to the condition
\beq p'(n^i-n^{N-i})=p(m^i-m^{N-i})\eeq
but there are no solutions to this for $ m,\,n$ in the range (\ref{mnrange}) 
if $p',\,p$ are relatively prime.

\item $\alpha^*\notin\{\alpha_w\}$. This implies that $\lambda(m,n)$
 cannot be the weight of a self conjugate representation. There are thus 
\emph{two} primary fields which are conjugates of each other,
 one built on $\lambda(m,n)$ and one
 on $-\lambda(m,n)$.
\end{enumerate}

The vertex operators $V_\alpha$ operating on the SL(2,C) invariant vacuum
generate states in a Fock space $\C F(\lambda(m,n))$ 
(where $\alpha$ and $\lambda$ are related as in in (\ref{alphadef})),
 rather than the Verma module
of the Virasoro primary field. This physical Hilbert space
 has to be constructed by a BRST procedure that was first described by Felder
\cite{Felder:1989zp} and extended to the $W_3$ case in \cite{Mizoguchi:1992vt}. 
First define the set of operators
\beq Q^{(j)}_k=B^j_k\left(\oint dz V_{\alpha_+e_j}(z)\right)^k,
\quad j=1\ldots r,\; k<p,\eeq
where the $B^j_k$ are non-zero constants (note that $h(\alpha_\pm e_i)=1$). The $ Q^{(j)}_k$
 commute with the Virasoro algebra by construction and map
\beq Q^{(j)}_k\C F(\lambda)\to\C F(\lambda-kp'e_j)\eeq
It is simple to check that if
\beq k=n^j\;\hbox{mod}\; p\label{kcond}\eeq
then the conformal dimensions of these two Fock spaces differ by an integer.
The action on $\lambda$ then amounts to
\beq  Q^{(j)}: \lambda(m,n)\to
-pm^i\omega_i+p'w_{e_j}\, n^i\omega_i - Npp'e_j\eeq
where $N\in\mathbb Z$ is   introduced to enforce the mod $p$ condition in
(\ref{kcond}). Now $Q^{(j)}$ and $Q^{(j+1)}$ do not commute so we have to
introduce further operators $Q^{(j,j+1)}$ such that 
\beq Q^{(j)}_{n^j}Q^{(j+1)}_{n^{j+1}}=Q^{(j,j+1)}_{n^j+n^{j+1}}Q^{(j)}_{n^j}\eeq
and with action on $\lambda$,
\beq  Q^{(j,j+1)}: \lambda(m,n)\to
-pm^i\omega_i+p'w_{e_{jj+1}}\, n^i\omega_i - Npp'e_{jj+1}.\eeq
This operator in turn does not commute with 
(eg) $Q^{(j+2)}$ and so we iterate this process
 ending up with a set of operators $Q^{(j,k)}$ with action on $\lambda$ given by
\beq  Q^{(j,k)}: \lambda(m,n)\to
-pm^i\omega_i+p'w_{e_{jk}}\, n^i\omega_i - Npp'e_{jk}.\eeq

Starting from a given Fock space chosen according to the rules 
 (\ref{alphadef}) one can now convince oneself that 
the action of the $Q^{j,k}$ generates an infinite complex
of Fock spaces\footnote{This is hard to draw unless $r=2$ for which 
case it is described in detail in \cite{Mizoguchi:1992vt,Caldeira:2003zz}}
\beq \C C(\lambda)=
\bigoplus_{  \substack{w\in {\mathcal W}\\ N\in {\mathbb Z}^r}  }
\C F(-pm^i\omega_i+p'w\, n^i\omega_i - pp'N^ie_i).\eeq
It is possible to assemble from the $Q^{j,k}$
 a nil-potent operator $Q_B$ on $\C C(\lambda)$
whose cohomology is the physical Hilbert space \footnote{This was
 proved by Felder for the $r=1$ case; there seems to be no proof 
given in the literature of
the more general case,  but there is also no
evidence to the contrary.}
\beq \C H=\frac{\mathrm{Kernel}\, Q_B}{\mathrm{Image}\, Q_B}.\label{coho}\eeq
Expectation values are then calculated from
  alternating sums  over the complex so for example the
 character of the Verma module is given by
\bea \chi_{\lambda(m,n)}(q)&=&\mathrm{Tr}\, q^{L_0-c/24}\nn\\
&=&\frac{1}{\eta(\tau)^{r}}
\sum_{  \substack{w\in {\mathcal W}\\ N\in {\mathbb Z}^r}  }
\varepsilon_w q^{|p^{\prime}w n^i\omega_i-pm^i\omega_i
+pp^{\prime}N^ie_{i}|^{2}/2pp'}.
\label{character}
\eea

\section{Coherent boundary states\label{COHSTATES}}

Coherent boundary states may be defined in a straightforward 
generalization of the procedure for the one component Coulomb gas
\cite{Kawai:2002vd,Caldeira:2003zz}. First we introduce
the states $|\alpha,\bar\alpha\rangle$ which are 
 constructed by applying the vertex operator
$V_{\alpha}(z)$ and its antiholomorphic counterpart, ${\overline
V}_{\bar\alpha}(\overline{z})$
 to the
$SL(2,C)$-invariant vacuum $|0,0\rangle$,
\begin{equation}
|\alpha,\bar{\alpha}
\rangle=\lim_{z,\overline{z}\rightarrow0}\overline V_{\bar\alpha}(\overline
{z})V_{\alpha}(z)|0,0\rangle=e^{i\overline{\alpha}
\cdot\overline \phi_{0}}e^{i\alpha\cdot\phi_{0}}|0,0\rangle.
\end{equation}
These states satisfy
\begin{eqnarray}
 a_{0}^{i}|\alpha,\bar{\alpha}
\rangle&=&\alpha^{i}|\alpha,\bar{\alpha}\rangle,\nonumber\\
\bar{a}_{0}^i|\alpha,\bar{\alpha}\rangle&=&\bar{\alpha}^{i}|\alpha,\bar{\alpha}\rangle.
\end{eqnarray}
The corresponding bra states are given by
\begin{equation}
\langle\alpha,\bar{\alpha}|
=\langle 0,0| e^{-i\overline{\alpha}
\cdot\overline \phi_{0}}e^{-i\alpha\cdot\phi_{0}}.
\end{equation}
The  coherent  state ansatz is given by
\bea
|B(\alpha,\bar\alpha;\Lambda)\rangle&=&C_\Lambda
\,|\alpha,\bar{\alpha}\rangle,\\
C_\Lambda&=&\prod_{k>0}\exp\left(\frac{1}{k}\,a_{-k}
\cdot\Lambda\cdot \bar{a}_{-k}\right) ,
\eea
where $\Lambda$ is a matrix to be determined by imposing the boundary condition
\begin{equation}
(L_{n}-\bar{L}_{-n})|B(\alpha,\bar{\alpha};\Lambda)\rangle=0.\label{LLbar}
\end{equation}
For positive $n$ this gives the constraint
\bea
\Big(&&\half\sum_{l=1}^{n-1}\abar_{n-l}\cdot(\Lambda^T\cdot\Lambda-I)\cdot\abar_{-l}
+(a_0-2\alpha_0(n+1)\rho)\cdot\Lambda\cdot\abar_{-n}\nn\\
&&+(-\abar_0-2\alpha_0(n-1)\rho)\cdot\abar_{-n}\Big) \ket{\alpha,\bar{\alpha}}
=0,\label{lambdacond}%
\eea
 and  similarly  for negative $n$. The constraint is satisfied provided
\bea \Lambda^T\cdot\Lambda&=&I,\label{Lortho}\\
\Lambda\cdot\rho+\rho&=&0,\label{Levect}\\
\Lambda^T\cdot\alpha+4\alpha_0\rho-\bar{\alpha}&=&0.\label{lambdacondA}
\eea
The last of these conditions allows us to simplify our notation by defining
\beq
|B(\alpha;\Lambda)\rangle\equiv|B(\alpha,\bar{\alpha}=\Lambda^T\cdot\alpha+4\alpha_0\rho;\Lambda)\rangle.\label{Bdef}\eeq

We next identify solutions for $\Lambda$.
Using the form (\ref{alphadef}),
%\beq \alpha(F)=2\alpha_0\rho -\alpha_- m^i\omega_i
%-\alpha_+ n^i\omega_i
%\eeq
%and similarly for $\bar\alpha$
 the last constraint in (\ref{lambdacondA})
 becomes
\beq \Lambda^T (p n^i-p'm^i)\omega_i=-(p \bar n^i-p'\bar m^i)\omega_i.\eeq
Using the $\{\omega_i\}$ basis, we see that the simplest form for $\Lambda$,
which we will denote $\Lambda^\omega$,
is one in which all the elements are integers; this guarantees that $\alphabar$
exists. In addition  the vector $(1,\ldots 1)$ must be an eigenvector of
 $\Lambda^\omega$ with eigenvalue $-1$ in order that (\ref{Levect})
 is satisfied
so 
\beq \Lambda^\omega_{k1}+\Lambda^\omega_{k2}+\ldots+\Lambda^\omega_{kr}=-1,
\quad k=1,\ldots r\label{Lsum}.\eeq
 A little bit of care is necessary in implementing (\ref{Lortho});
 recall that this
is in an \emph{orthogonal cartesian} basis because of the definition of
 the Heisenberg algebra (\ref{heisenberg}). In the  $\{\omega_i\}$ basis it
 becomes
\bea \Lambda^\omega\,^T A^{-1}\, \Lambda^\omega&=&A^{-1},\quad
\hbox{or, equivalently,\hfill}\quad
\Lambda^\omega\, A\, \Lambda^\omega\,^T=A\label{LAL},\eea
where $A$ is the Cartan matrix; picking out the diagonal elements gives
\beq (\Lambda^\omega_{k1})^2+(\Lambda^\omega_{k1}-\Lambda^\omega_{k2})^2+
\ldots +(\Lambda^\omega_{k\,r-1}-\Lambda^\omega_{kr})^2+
(\Lambda^\omega_{kr})^2=2.\label{LALsum}\eeq
Now (\ref{LALsum}) is a sum  of squares so exactly two terms in the sum must
be equal to unity. In conjunction with (\ref{Lsum}) this shows that 
 each row of $\Lambda^\omega$ contains one element which is $-1$, all
 other elements being zero. Each row must be different, otherwise
 $\det\Lambda^\omega=0$ which contradicts  (\ref{LAL}) because $\det A\ne 0$.
Thus the action of $\Lambda^\omega$ on the l.h.s. of (\ref{LAL}) is to permute 
the rows and columns of $A$.  By inspection there are only two permutations 
that leave $A$ invariant, the identity and reversal
 of the order of rows and columns, so  there are only two
 solutions 
\bea \Lambda^\omega&=&-I\nn\\
\hbox{or}\quad \Lambda^\omega&=&S.\eea
Equivalently, the action of $\Lambda$ is simply the group
 of outer automorphisms on the Dynkin diagram for $A_r$.

\section{States and decoupled states\label{COUPLED}}

The cylinder amplitudes between boundary states of the form (\ref{Bdef}) 
can be calculated by standard techniques
 and are given by
\beq \bra{B(\beta;\Lambda_2)}q^{\Ham}\ket{B(\alpha;\Lambda_1)}=
q^{h(\alpha)-\cbytwfo}\exp\left(\sum_{k=1}^\infty\sum_{l=1}^\infty
\frac{ q^{kl}}{l}\Tr(\Lambda_1\Lambda_2^T)^l\right)\delta_{\alpha,\beta}\,
\delta_{\alpha,\Lambda_1\Lambda_2^T\beta}.\label{cylamp}
\eeq
There are then three cases where the amplitude is non-zero;
\bea \bra{B(\alpha;\Lambda_1)}q^{\Ham}\ket{B(\alpha;\Lambda_2)}&=&\frac 
{q^{h(\alpha)-\cbytwfo}}{\prod_{k>0}(1-q^k)^r},\quad \Lambda_1=\Lambda_2,
 \nn\\
&=&\frac {q^{h(\alpha)-\cbytwfo}\, \delta_{\alpha,-S\alpha}}
{\prod_{k>0}(1-q^{2k})^{\frac r 2}},
\quad \Lambda_1\ne\Lambda_2,\quad r\hbox{~even} \nn\\
&=&\frac {q^{h(\alpha)-\cbytwfo}\, \delta_{\alpha,-S\alpha}}
{\prod_{k>0}(1-q^k)(1-q^{2k})^{\frac {r-1} 2}},
\quad \Lambda_1\ne\Lambda_2,\quad r\hbox{~odd}.\nn\\
\left.\right. 
\eea
Note that  the states $\ket {B(\alpha;S)}$ can be written down for any 
$\alpha$ but they are completely decoupled from the theory unless
\beq \alpha=-S\alpha\label{alpconstraint}\eeq
and, from (\ref{lambdacondA}), all non-decoupled states have the property
\beq \alpha+\alphabar=4\alpha_0\rho.\label{alpsym}\eeq
The constraint (\ref{alpconstraint}) implies that
\beq \lambda(m,n)=-S\lambda(m,n)\eeq
which is uniquely satisfied by the highest (or lowest) weights of 
self-conjugate representations. Thus  the only primary fields 
which have the second boundary state $\ket {B(\alpha;S)}$
associated with them are the self-conjugate ones.

 The states $\ket {B(\alpha;\Lambda)}$ lie in the Fock space and the
corresponding  states that lie in the physical Hilbert space,
 $\ishiket {\alpha;\Lambda}$, are obtained by summing
over the Felder complex
\beq \ishiket {\alpha;\Lambda}=
\sum_{  \substack{w\in {\mathcal W}\\ N\in {\mathbb Z}^r}  }
\kappa^{\phantom\prime}_{wN}\ket {B(2\alpha_0\rho-\frac{1}{\sqrt{pp^\prime}}\,
(p^{\prime}w n^i\omega_i-pm^i\omega_i
+pp^{\prime}N^ie_{i});\Lambda)}\label{ishistate}\eeq
where the $\kappa_{wN}$ are constants of magnitude 1. There is a similar 
expression for the bra states but with 
 $\kappa^{\phantom\prime}_{wN}$ replaced by
$\kappa'_{wN}$ satisfying
\beq \kappa^{\phantom\prime}_{wN}\kappa_{wN}'=\varepsilon_w.\eeq
From these states the physical cylinder amplitudes can be calculated;
between identical in and out states these are simply the characters
\beq \ishibra{\alpha;\Lambda}q^{\Ham}\ishiket{\alpha';\Lambda}
=\chi_{\lambda(m,n)}(q)\delta_{\alpha,\alpha'}.\eeq
However for the self-conjugate fields there
 is a second non-zero amplitude 
\beq \widetilde\chi_{\lambda(m,n)}(q)=
\ishibra{\alpha;-I}q^{\Ham}\ishiket{\alpha;S}.\label{mixeda}\eeq
Recall that, for such fields, $n^i\omega_i$ and $m^i\omega_i$ must
 be self-conjugate highest weights;  non-zero contributions to this amplitude
further require that $w n^i\omega_i$ and $N^ie_{i}$ are self-conjugate.
To solve these constraints introduce the basis for 
self-conjugate combinations of roots
\beq d_k=e_k+e_{k+1}+\ldots+e_{N-k},\quad k=1,\ldots
\textstyle{ \lceil \frac r 2\rceil}\label{dbasis}
\eeq
which, conveniently, is also orthogonal and denote by
$\widetilde{ \C W}$ the abelian subgroup of $\C W$ 
\beq \bigotimes_{k=1}^{ \lceil \frac r 2\rceil}\{I,w_{d_k}\}.\eeq
Then the amplitude becomes
\bea \widetilde\chi_{\lambda(m,n)}(q)
&=&\frac{1}{\eta(\tau)^{r-2\lfloor \frac r 2\rfloor}\eta(2\tau)^{\lfloor
 \frac r 2\rfloor}}
\sum_{  \substack{w\in\widetilde{ \C W}
 \\ N\in {\mathbb Z}^{\lceil \frac r 2\rceil}  }}
\varepsilon_w q^{|p^{\prime}w n^i\omega_i-pm^i\omega_i
+pp^{\prime}N^id_{i}|^{2}/2pp'}.
\label{mixedamp}
\eea
This expression can be simplified further but it is better  to 
study the modular properties of the diagonal cylinder amplitudes first.

\section{Modular properties\label{MODPROPS}}

The modular properties of the cylinder amplitudes can be 
examined using standard methods. Starting with the diagonal amplitudes,
or equivalently the characters (\ref{character}), the Poisson resummation 
formula gives
\bea \chi_{\lambda(m,n)}(q)
&=&\frac{1}{(pp')^{\half r} \eta(\tau')^{r}\sqrt{\det A}}
\sum_{  \substack{w\in {\mathcal W}\\ \widetilde N\in {\mathbb Z}^r}  }
\varepsilon_w q'^{|\widetilde N^i\omega_i|^2/2pp}
\,e^{i2\pi (\widetilde N^i\omega_i)\cdot( p^{\prime}w n^i\omega_i-pm^i\omega_i)
/pp'},
\label{modchar}
\eea
where $\tau'=-1/\tau$ and $q'=e^{i2\pi\tau'}$. Now split the sum over the
 dual lattice 
\beq\C D= \mathbb Z \omega_1+\ldots+\mathbb Z \omega_r\eeq
into a sum over the scaled root lattice
\beq\C R= pp'(\mathbb Z e_1+\ldots+\mathbb Z e_r)\eeq
and the quotient $\C Q=\C D/\C R$\footnote{There are many equivalent choices
for $\C Q$; in the following we take it to be the interior of a polyhedron
centred on the origin.}.
Then it is the case that
\beq \widetilde N^i\omega_i=pp'N^ie_i + b,\eeq
where $N\in\mathbb Z^r$ and  $b=b^i\omega_i\in \C Q$
 and (\ref{modchar}) becomes
\bea \chi_{\lambda(m,n)}(q)
&=&\frac{1}{(pp')^{\half r} \eta(\tau')^{r}\sqrt{\det A}}
\sum_{  \substack{w\in {\mathcal W}\\  N\in {\mathbb Z}^r\\b\in \C Q} }
\varepsilon_w q'^{|  pp'N^ie_i + b  |^2/2pp'}
\,e^{i2\pi  b\cdot( p^{\prime}w n^i\omega_i-pm^i\omega_i)
/pp'}.
\label{emodchar}
\eea

All  points
in $\C Q$ can be written
$b=p'k^i\omega_i-p\ell^i\omega_i$. It
is therefore convenient to introduce the unique integers $r_0,s_0$ such that
\bea 1&\le&r_0\le p-1,\nn\\
1&\le&s_0\le p'-1,\nn\\
1&=&p'r_0-ps_0,\eea
and define the operators
\bea P_w&=&p'r_0w-ps_0,\nn\\
\overline P_w&=&w P_{w^{-1}}=  p'r_0-ps_0w,\eea
which have the properties
\bea P_w (p^{\prime} k^i\omega_i-p\ell^i\omega_i)
&=&p^{\prime}w\, k^i\omega_i-p\ell^i\omega_i\quad\hbox{mod }\C R,\nn\\
\overline P_w (p^{\prime} k^i\omega_i-p\ell^i\omega_i)
&=&p^{\prime} k^i\omega_i-p w\, \ell^i\omega_i\quad\hbox{mod }\C R,\nn\\
 P_w P_{w'}&=&P_{ww'}\quad\hbox{mod }\C R,\nn\\
 \overline P_w\overline P_{w'}&=&\overline P_{ww'}
\quad\hbox{mod }\C R.\label{comboC}\label{comboB}\eea
Note that $P_w$ is defined so that repeated application of it simply
generates the Felder complex $\C C(\lambda)$.

Now,  suppose that 
\beq b=P_{\tilde w} b\;\hbox{mod }\C R,\label{Pcond}\eeq where $\tilde w$ is
 some odd element
of the Weyl group;
then $b$ can  be replaced by $P_{\tilde w}b$ in the phase factor
part of (\ref{emodchar}); using (\ref{comboB}) and
changing the summation variable over  $\C W$ to $w \tilde w$
shows that 
the contribution to $\chi$ is minus itself and therefore must be zero.
A similar argument applies if $b=\overline P_{\tilde w} b\;\hbox{mod }\C R$.
The condition \ref{Pcond} implies that
\beq \tilde w\, k^i\omega_i=k^i\omega_i+pN^ie_i.\label{PcondA}\eeq
Choosing $\tilde w$ to be a reflection in an arbitrary root
 $e_{m,n}=$ shows that (\ref{Pcond}) is certainly 
 the case if
\beq \sum_{i=m}^n k_i = 0\quad\hbox{ mod $p$, for some
$m,n$: $r\ge n\ge m >0$}.\label{kkcond}\eeq
A similar exercise on $\overline P$ yields the conditions
\beq \sum_{i=m}^n \ell_i = 0\quad\hbox{ mod $p'$, for some
$m,n$: $r\ge n\ge m >0$}.\eeq 
It follows that only 
those $b$s for which  
\beq \sum_{i=m}^n b_i \ne 0 \quad\hbox{mod $p'$ \emph{or} mod $p$, for any
$m,n$: $r\ge n\ge m >0$}.\label{bcond}\eeq
contribute to the sum over $\C Q$ in (\ref{emodchar}).
In particular for the $b$s remaining $b^i,k^i,\ell^i\ne0$ and so 
$b$, $k^i\omega_i$ and $\ell^i\omega_i$ all lie inside Weyl chambers
(never on the boundaries) and can always be moved to the fundamental
 Weyl chamber $\C C_0$ by the application of a Weyl transformation. Then if
$b$ lies in $\C C_0\cap\C Q$
\beq b\cdot\theta\le \frac{pp'}{2}\theta\cdot\theta\eeq
ie $\sum_{i=1}^rb^i\le pp'$; but equality is ruled out by (\ref{bcond}) so
$b$ always lies inside $\C Q$ (never on the boundary). 

Now consider $k^i\omega_i$ in $\C C_0$ and define the hyperplanes $\Pi_M$ by
\beq \Pi_M:\quad (x-\half Mp\theta)\cdot \theta=0,\quad M=1,2,\ldots .\eeq
Note that because of (\ref{kkcond}) $x=k^i\omega_i$ can never lie
 on the hyperplanes; supposing that $x$ lies between $\Pi_M$ and $\Pi_{M+1}$
reflect it in $\Pi_M$ to get 
\beq x'=w_\theta x+Mp\theta.\eeq
If $x'$ lies outside $C_0$ a Weyl transformation will put it back so
\beq T x=w(w_\theta x+Mp\theta)\eeq
lies between  $\Pi_{M-1}$ and $\Pi_{M}$ and in $\C C_0$.
 Successive transformations will shift
$x$ to the region between $\Pi_{0}$ and $\Pi_{1}$ and in $\C C_0$;
 in this region
$x$ satisfies
\beq x\cdot\theta< p.\eeq
The effect of $T$ on $b$ is
\bea b&\to& p'w(w_\theta k^i\omega_i +Mp\theta)-p\ell^i\omega_i\nn\\
&=&P_{ww_\theta}b\quad \hbox{mod $\C R$.}\eea
Similar manipulations on $\ell^i\omega_i$ lead to the conclusion that any
$b$ can be written in the form
\bea b&=&w'P_w b_0,\nn\\
b_0&=&p'k^i\omega_i-p \ell^i\omega_i, \eea
where $k^i $ and $\ell^i$ 
are positive
integers satisfying
\beq \sum_{i=1}^r\ell^i<p',\quad\sum_{i=1}^rk^i<p.\label{klrange}\eeq

The $b_0$s are the same as the set of allowed $\lambda$s derived 
from fusion and discussed in Section \ref{WMIN}. However we have not finished. 
Clearly some $b_0$s fall into $\C C_0$; these satisfy
\bea \theta\cdot b_0=\sum_{i=1}^r b_0^i&=&
p'\sum_{i=1}^r k^i-p\sum_{i=1}^r\ell^i\nn\\
&\le&p'(p-1)-rp.\eea
Others do not  fall into $\C C_0$ but can be put there by Weyl transformation,
$b_0'=w b_0$,  so that 
\beq \theta\cdot b_0'=\pm \sum_{i=m}^n p'k^i-p \ell^i\quad\hbox{for some
$m,n$: $r\ge n\ge m >0$,}\label{prev}\eeq
where we have used the fact that $w\theta$ is a root. Thus we get the bounds
\bea \sum_{i=1}^r  {b_0'}^i&\le& p'(p-1)-rp -(r-n+m-1)(p'-p),\quad\hbox{if 
``$+$'' in (\ref{prev}),}\nn\\
&\le& p'(p-1)-rp -(n-m)(p'-p),\quad\hbox{if ``$-$'' in (\ref{prev}),}\eea
and so the $b_0$s are equivalent, up to a Weyl transformation, to a sub-set of
$b_+$s defined by
\beq b_+^i\ge 1,\quad \sum_{i=1}^r  {b^i_+}\le p'(p-1)-rp,\eeq
and the conditions (\ref{bcond}). 
The transformations $T$ are invertible so there are no $b$s which cannot be
 obtained by starting with a point outside the $b_+$ domain and applying T.
 However some points in the $b_+$ domain may correspond to either $k^i\omega_i$
or
 $\ell^i\omega_i$ lying outside (\ref{klrange}). (The possibility of
cancellation
 between the two terms in $b$ ensures that it is never the situation that
\emph{both} lie outside.)
Suppose that $k^i\omega_i$ lies outside then for some $m,n$ 
\bea \sum_m^n{b^i_+}&=&(p+K)p'-p(p'-L),\quad 0<K,\; 0<L<p\nn\\
&=& Kp'+Lp,\quad 0<K<p,\;0<L<p'.\eea
A similar argument deals with the case that $\ell^i\omega_i$ lies outside,
except
 that this time we must take ``$-$'' in (\ref{prev}).
Thus if
\beq \sum_m^n{b^i_+}=Kp'+Lp,\quad 0<K<p,\;0<L<p',\eeq
then $b_+$ cannot be one of the $b_0$s.
We conclude that all $b$s
 contributing to the sum over $\C Q$ in (\ref{emodchar}) can be written
\beq b=w'P_w\mu,\quad w',w\in \C W,\label{mudef}\eeq
where $\mu=\mu^i\omega_i$ lies in the set $\C B$ defined by
\bea \C B:\qquad \mu^i&\ge&1,\nn\\
 \sum_{i=1}^r  {\mu^i}&\le& p'(p-1)-rp,\nn\\
\sum_m^n \mu^i&\ne&Kp'+Lp,\quad 0<K<p,\;0<L<p',\nn\\
&&\qquad\hbox{ for any
$m,n$: $r\ge n\ge m >0$}.\label{PFrules}\eea
It is also convenient to define the self-conjugate subset
\beq \C C:\quad \mu\in\C B;\,\mu=-S\mu.\eeq
Each $\mu\in\C B$ labels a primary field in the CFT and each
 distinct primary field has just one $\mu$. For example, in the case
 %in the $r=2$, $p'=5$, $p=4$, case 
$W_3(5,4)$ (the critical point of the
 3-state Potts model) application of the rules (\ref{PFrules}) yields
\bea I:\;\mu&=&\omega_1+\omega_2\nn\\
 \sigma:\;\mu&=&2\omega_1+\omega_2\nn\\
 \sigma^\dagger:\;\mu&=&\omega_1+2\omega_2\nn\\
\psi:\;\mu&=&6\omega_1+\omega_2\nn\\
 \psi^\dagger:\;\mu&=&\omega_1+6\omega_2\nn\\
\epsilon:\;\mu&=&3\omega_1+3\omega_2,\eea
whereas  $W_4(5,4)$ yields just one solution
\beq \mu=\omega_1+\omega_2+\omega_3,\eeq
which is the identity operator, and so on.

The characters $\{\chi_\mu(q), \;\mu\in \C B\}$  form a representation of the
modular group. Combining (\ref{mudef}) and (\ref{emodchar}) gives
\bea  \!\!\!\! \!\!\!\! \!\!\!\! \!\!\!\!  \chi_\lambda(q)
&=&\frac{1}{(pp')^{\half r} \eta(\tau')^{r}\sqrt{\det A}}
\sum_{  \substack{w, w',w''\in {\mathcal W}\\  N\in {\mathbb Z}^r\\ \mu\in \C B
}}
\varepsilon_w q'^{|  pp'N^ie_i + w''P_{w'}\mu  |^2/2pp'}
\,e^{i2\pi w''P_{w'} \mu\cdot P_{w}\lambda
/pp'},
\label{eemodchar}
\eea
which can be simplified by changing variables to $\{w'',w',w'w\}$ and
using (\ref{comboC}) to
\bea \!\!\!\! \!\!\!\! \!\!\!\! \!\!\!\! \chi_\lambda(q)
&=&\frac{1}{(pp')^{\half r} \eta(\tau')^{r}\sqrt{\det A}}
\sum_{  \substack{w, w',w''\in {\mathcal W}\\  N\in {\mathbb Z}^r \\ \mu\in \C B
}}
\varepsilon_w\varepsilon_{w'} q'^{|  pp'N^ie_i + P_{w'}\mu  |^2/2pp'}
\,e^{i2\pi w'' \mu\cdot P_{w}\lambda
/pp'},
\eea
which is nothing but
\beq \chi_\lambda(q)=\sum_{\mu \in \C B} S_{\lambda \mu}   \chi_\mu(q'),\eeq
where
\beq S_{\lambda \mu}=\frac{1}{(pp')^{\half r} \sqrt{\det A}}
\sum_{w,w'\in\C W}{\varepsilon_{w}}
\,e^{i2\pi \mu\cdot w' P_w\lambda
/pp'}.\label{Fexp}
\eeq
Thus we see that the characters for the set of primary fields given
by (\ref{PFrules}) are covariant under modular transformations. 
Note that $S$ is symmetric, real if both $\lambda,\mu\in\C C$, and that
$S_{I\lambda}>0$ by Cardy's argument \cite{Cardy:1989ir}.
One 
can easily check by explicit evaluation that (\ref{Fexp}) generates the correct
modular $S$ matrix for the three state Potts model. It is easy to check that
$S$ is unitary and that $S^2=C$, the charge conjugation matrix; for example
\bea S^2_{\lambda\rho}&=&\frac{1}{(pp')^{ r} {\det A}}\sum_{\mu\in\C B}
\sum_{w,w'\in\C W}\sum_{\bar w,\bar w'\in\C W}{\varepsilon_{w}}
{\varepsilon_{\bar w}}\,e^{i2\pi
 \mu\cdot( w' P_w\lambda+\bar w' P_{\bar w}\rho)
/pp'}.\eea
Now change variables to $b=w'P_w\mu$ and reintroduce those 
$b$s which were excluded in (\ref{bcond})
because they did not contribute to the sum;
 they don't contribute here either so the sum over $w,w',\mu$
 can be replaced by the unrestricted sum over $\C Q$ to give, after 
a change of variables in the remaining Weyl sum,
\bea S^2_{\lambda\rho}&=&\frac{1}{(pp')^{ r} {\det A}}\sum_{b\in\C Q}
\sum_{ w, w'\in\C W}{\varepsilon_{w}}
\,e^{i2\pi
 b\cdot( \lambda+ w' P_{ w}\rho)
/pp'}.\eea
The sum over $\C Q$ is now zero unless $\lambda+ w' P_{ w}\rho=0$  which occurs
only if $w=1$, $w'=w_0$ and $\rho=-w_0\lambda$, ie $\rho$ and $\lambda$ are
conjugate representations; in which case the sum just gives
 the volume of $\C Q$ which cancels the denominator and leaves
$S^2_{\lambda\rho}=1$. A similar argument can be used to demonstrate unitarity.

We now return to the mixed amplitudes \ref{mixedamp}
which can be rewritten using the notation of this section as
\bea \widetilde\chi_{\lambda}(q)
&=&\frac{1}{\eta(\tau)^{r-2\lfloor \frac r 2\rfloor}\eta(2\tau)^{\lfloor
 \frac r 2\rfloor}}
\sum_{  \substack{w\in\widetilde{ \C W}
 \\ N\in {\mathbb Z}^{\lceil \frac r 2\rceil}  }}
\varepsilon_w q^{|pp^{\prime}N^id_{i}+P_w\lambda|^{2}/2pp'},
\label{mixedampA}
\eea
where $\lambda\in\C C$. Using the basis
\ref{dbasis} in which
\beq \lambda=\half\hat\lambda_i d_i,\quad \hat\lambda_i=\lambda\cdot d_i,\eeq
$\widetilde\chi_{\lambda}(q)$ takes the factorized form
\bea \widetilde\chi_{\lambda}(q)&=&\prod_{i=1}^{\half r}\widehat
\chi_{\hat\lambda_i}(q),\quad\hbox{if $r$ even,}\nn\\
&=&\chi^V_{\hat\lambda_{\frac{r+1}{2}}}(q)\prod_{i=1}^{\half (r-1)}\widehat
\chi_{\hat\lambda_{i}}(q),\quad\hbox{if $r$ odd,}
\eea
where
\bea\widehat \chi_{\hat\lambda}(q)&=&\frac{1}{\eta(2\tau)}
\sum_{  \substack{\varepsilon=\pm 1
 \\ N\in {\mathbb Z}}}\varepsilon 
q^{pp^{\prime}(N+P_\varepsilon  \hat\lambda/2pp')^2},\nn\\
 \chi^V_{\hat\lambda}(q)&=&\frac{1}{\eta(\tau)}
\sum_{  \substack{\varepsilon=\pm 1
 \\ N\in {\mathbb Z}}}\varepsilon 
q^{pp^{\prime}(N+P_\varepsilon  \hat\lambda/2pp')^2},
\eea
and
\beq P_\varepsilon=p'r_0\varepsilon-ps_0.\eeq
The behaviour of these functions under modular transformation, which will be
needed in the next section, is given by
\bea \widehat \chi_{\hat\lambda}(q)&=&\frac{2}{\sqrt{pp'}}\sum_{a=1}^{ pp'-1}
\sum_{\varepsilon=\pm 1}\varepsilon\cos\left(\frac{\pi a
P_\varepsilon\hat\lambda}{pp'}\right)
G_{a}(q'),\nn\\
G_{a}(q)&=&\frac{1}{\eta(\tau/2)}\sum_{N\in {\mathbb Z}}
q^{pp^{\prime}(N+ a/2pp')^2},\label{modtwa}
\eea
and
\bea \widehat \chi^V_{\hat\lambda}(q)&=&\frac{\sqrt{2}}{\sqrt{pp'}}
\sum_{a=1}^{ pp'-1}\sum_{\varepsilon=\pm 1}\varepsilon\cos\left(\frac{\pi a
P_\varepsilon\hat\lambda}{pp'}\right)
G^V_{a}(q'),\nn\\
G^V_{a}(q)&=&\frac{1}{\eta(\tau)}\sum_{N\in {\mathbb Z}}
q^{pp^{\prime}(N+ a/2pp')^2}.
\eea
In these formulae the sum over $a$ can  omit  multiples of $p'$ and $p$
because the coefficient vanishes.
 Note that the set of functions appearing on the right hand is now not 
 the same as on the left hand side and at this point there is no particular 
relationship between the indices $\hat\lambda$ and $a$.
 Using these formulae shows that 
\beq \widetilde\chi_\lambda(q)=\sum_{\psi\in \C P}
 \Psi_{\lambda\psi} H_\psi(q'),\eeq
where $H_\psi(q')$ are some set of modular functions
 assembled from products of the $G$s,
and $\C P$ is some domain not yet determined but
 which contains at least as many
members as $\C C$.

\section{Physical boundary states\label{PHYSICAL}}

Physical boundary states are defined so that when
cylinder amplitudes between them are expressed in terms of the annulus 
variable $q'$ the result is a power series in $q'$ in which every coefficient
is a (positive) integer. Thus the annulus partition function with
 given boundary conditions
is essentially formed from the contributions of physical degrees
 of freedom propagating round the
annulus.
The $W_3(5,4)$ (critical three state Potts) model is also a Virasoro
 minimal model which can
 be exploited to ease the calculation of the physical boundary
 states. In general these shortcuts
are not available and we have to proceed rather differently.

 First  introduce the condensed
 notation  for the basis states in the Hilbert space (\ref{ishistate})
\bea \ishiket\lambda &\equiv& \ishiket{\alpha=2\alpha_0\rho-(pp')^{-\half}
\lambda,-I}\nn\\
\ishiket{\hat\lambda} &\equiv& \ishiket{\alpha=2\alpha_0\rho-(pp')^{-\half}
\lambda,S}.\eea
For every primary field there is an ordinary, $W$ current conserving, physical
 boundary state constructed in the usual
way \cite{Cardy:1989ir},
\beq \ket{\widetilde\lambda}= \sum_{\mu\in\C
B}\frac{S_{\lambda^*\mu}}{\sqrt{S_{I\mu}}}\ishiket\mu,\eeq
where $\lambda^*$ denotes the conjugate to $\lambda$.
In addition we expect there might
 be physical boundary states constructed from the $W$ violating states
\beq \ket{\widetilde A}=\sum_{\mu\in\C C} b_{A\mu}\ishiket{\hat\mu}.\eeq
 The annulus partition function with
boundary conditions labelled by $\widetilde\lambda$ and  $\widetilde A$ is then
\bea Z_{\widetilde\lambda\widetilde A}&=&\bra{\widetilde\lambda}q^{\Ham} 
\ket{\widetilde A}\nn\\
&=&\sum_{\mu\in\C C}\frac{S_{\lambda^*\mu}
b_{A\mu}}{\sqrt{S_{I\mu}}}\widetilde\chi_\mu(q)\nn\\
&=&\sum_{\mu\in\C C}\sum_{\psi\in \C P}\frac{S_{\lambda^*\mu}
b_{A\mu}}{\sqrt{S_{I\mu}}}
 \Psi_{\mu\psi} H_\psi(q').
\eea
Provided that the $H_\psi(q')$ are linearly independent the coefficient of
$H_\psi(q')$ should be an integer.
 The equations determining the boundary
states are then
\beq \sum_{\mu\in\C C}\frac{S_{\lambda^*\mu} b_{A\mu}}{\sqrt{S_{I\mu}}}
 \Psi_{\mu\psi}=n^\psi_{\widetilde\lambda\widetilde A}.\label{Bconds}\eeq

From now on we will confine our attention to the $(p+1,p)$ models which are
unitary
so there should be no question that the physical boundary states exist and
satisfy
the criteria discussed above.
$W_3(p+1,p)$ is
the  simplest case because there is only
 one basis vector, $d_1$ (\ref{dbasis}), and $pp'$ is necessarily even.
 Then (\ref{modtwa})
can be written
\bea \widetilde\chi_{\lambda}(q)&=&\frac{2}{\sqrt{pp'}}
\sum_{\mathrm{even~} a =2}
^{2pp'-2}\left(\cos\left(\frac{\pi(p'-p)a \hat\lambda}{2pp'}\right)-
\cos\left(\frac{\pi(p'+p)a \hat\lambda}{2pp'}\right)\right)
 G_{a/2}(q')\label{nasty1}\eea
Now regard the even number $a$ as being $\hat\mu=\mu\cdot d_1$ where $\mu$ is
a self conjugate vector. For the same reasons as in the previous section,
there is no contribution if $a$ is a multiple of $p$ or $p'$. Furthermore
$p'-p=1$  and multiplication by $p'+p$ mod $2pp'$
simply generates $a$ corresponding to the other self-conjugate member of
 the Felder orbit (\ref{mudef}). Thus all remaining $a$s in the
 sum correspond to $\mu\in\C C$, its partner in the orbit,
 or $2pp'$ minus one of these
(an explicit proof is given in Appendix A).
Multiplying a given  $a$ by $p'+p$ simply changes the overall
sign of the coefficient in (\ref{nasty1}). Taking all this into account
(\ref{nasty1})  becomes
\bea \widetilde\chi_{\lambda}(q)&=&\sum_{\mu\in\C C}\Psi_{\lambda\mu}
\left( G_{\hat\mu/2}(q') +G_{pp'-\hat\mu/2}(q')
 -G_{(p'+p)\hat\mu/2}(q') -G_{pp'-(p'+p)\hat\mu/2}(q')\right),\nn\\
\Psi_{\lambda\mu}&=&
\frac{4}{\sqrt{pp'}}
\sin\left(\frac{\pi\hat\mu \hat\lambda}{2p}\right)
\sin\left(\frac{\pi\hat\mu \hat\lambda}{2p'}\right).\label{fluke}\eea
The combination of $G$ functions appearing here has a series expansion in $q'$
with all positive coefficients and is linearly independent for different $\mu$.
Thus this is the form required; note that $\Psi$ is in fact a square matrix
 in this case and  it
is straightforward to check that 
\beq \Psi_{\lambda\rho}\Psi_{\rho\mu}=\delta_{\lambda\mu}.\eeq

When $p$ is small the conditions (\ref{Bconds}) can now be
 solved by brute force. For $p=4$ they yield exactly the ``free'' and ``new''
 boundary conditions found by Affleck et al \cite{Affleck:1998nq}.
To solve the constraints (\ref{Bconds}) one can proceed analogously to the 
calculation of the usual boundary states; suppose that 
$n^\psi_{\widetilde 0\widetilde A}=\delta^\psi_{\widetilde A}$ and use the
invertibility of $\Psi$ to get  
\beq  b_{A\mu}=\frac{ \Psi_{\mu A}}{\sqrt{S_{I\mu}}}.\eeq
 Substituting these back in (\ref{Bconds}) yields
\beq \sum_{\mu\in\C C} \frac{S_{\lambda^*\mu}
   \Psi_{\mu A}\Psi_{\mu \psi} } {{S_{I\mu}}}
 =  n^\psi_{\widetilde\lambda\widetilde A}. \eeq
For $p=5$, which has  four self-conjugate fields with $\hat\mu=2,4,8,14$,
 and
for $p=6$, which has six with $\hat\mu=2,4,8,10,16,22$, 
 we have checked explicitly that the expressions on the left hand
 side generate positive integers. 
Thus these models generate an analogue of the 
Verlinde formula for the symmetry violating sector 
(see also \cite{Fuchs:1999zi,Fuchs:1999xn} for other
examples of this).

For rank 3 and above the situation is more complicated and 
even for the simplest rank 3 model, $W_4(6,5)$, we have
not been able to construct
consistent physical states in the $W$ violating sector. This suggests that the
basis of
symmetry violating states may be incomplete (further evidence for this is given
in
the next section) and we will return to this problem in a separate paper.

\section{$W$ currents\label{WCURRENTS}}
In this section we will assume that the $W$ fields can indeed be constructed 
according to the prescription of \cite{Fateev:1988xx} and
 show that $\ket {B(\alpha;S)}$
necessarily violates conservation of \emph{all} the currents with conformal
 dimension greater than 2.
The first step in obtaining the Virasoro primary fields
 $\{W_K,\, K=3,\ldots N\}$
which, together with $T$, form  the $W$ algebra is to
define the generating 
functional 
\beq
(2i\alpha_{0})^{N}\mathcal{D}_{N}    =\,:\prod_{K=1}^{N}\left(  2i\alpha
_{0}\partial_{z}+h_{K}\cdot\partial\phi(z)\right):\,,\label{Ddef}
\eeq
where, if the product were written out, the value of $K$
 increases from left to right.
This can be evaluated \cite{Bilal:1991eu, Bouwknegt:1993wg} to get 
\begin{equation}
(2i\alpha_{0})^{N}\mathcal{D}_{N}=(2i\alpha_0\partial)^{N}+\sum_{K=1}^{N}
u_{K}[\phi(z)]\;
(2i\alpha_{0}\partial)^{N-K},
\end{equation}
where the $u_{K}[\phi(z)]$ are  fields
 of conformal dimension $K$ ($u_1$ vanishes identically). 
Unfortunately the $u_{K}$, apart from $u_2(z)\equiv T(z)$, are
 not themselves primary fields.
The true $W_K$ are constructed
from combinations of $\{u_K,\ldots, u_2,\partial\}$. In the present case it
 is sometimes
 more useful  to think of them being assembled iteratively as combinations
of $\{u_K,W_{K-1}\ldots, W_2,\partial\}$. Note that $W_K$ is essentially
just a normal ordered
multinomial expression in $\partial\phi^i$ and its derivatives but
 that unfortunately precise expressions are unknown for
 $K>5$ (for the completely  known algebras see \cite{Ozer:1998zg}). The
following  arguments will make use of the facts  that a)
\beq \C S_{K>1}(x) \equiv \sum_{L_1>L_2>\ldots L_K}(h_{L_1}\cdot x)
( h_{L_2}\cdot x)
\ldots (h_{L_K}\cdot x) =\frac{(-1)^{K+1}}{K}\sum_{L=1}^N ( h_{L}\cdot x)^K\eeq
 and b), defining $\C S_1(x)=1$,
\beq \sum_{k=1}^{ \lfloor{\scriptstyle\half} K\rfloor}
A_k \C S_{K-k}(x)\C S_k(x)\ne 0,\quad\hbox{unless}\, A_k=0,\forall k,
\label{Sindep}\eeq
(ie linear independence of the product functions).

Conservation of the $W$ current
 at the boundary is given in terms of modes by
\beq \left(W_{K\,n}-(-1)^K \Wbar_{K\,-n}\right)\ket {B(\alpha;\Lambda)}=0.\eeq
Since the $W_K$ are primary, it is sufficient to check the $n=0$ case so 
let
\beq \left(W_{K\,0}-(-1)^K \Wbar_{K\,0}\right)\ket {B(\alpha;\Lambda)}=
C_\Lambda\Delta_0^K\ket{\alpha,\alphabar},\eeq
and $\Delta_0^{KL}$ be the part of $ \Delta_0^K$ containing exactly
$L$ factors of $a_n$ or $\abar_n$ with $n\ne 0$.
Terms in $W_K$ which do not contain $\partial$ but just products of primaries
satisfy this condition automatically if the primaries do; if they do not
 then the violations cannot be cancelled between different products
 of primaries
on account of (\ref{Sindep}) and $W_K$ automatically violates
 the condition too.   So, of the
 possible terms in $W_K$ which contribute to $\Delta_0^{KK}$
 only the ones coming from $u_K$ have
 unknown properties at the boundary. Now note that
\bea h_K\cdot\partial\phi\, C_\Lambda\ket{\alpha,\alphabar}& =& -i C_\Lambda
\left(\frac{h_K\cdot \alpha}{z} 
+\sum_{n<0}\frac{h_K\cdot a_n}{z^{n+1}}+\sum_{n<0}\frac{h_K\cdot\Lambda
 \abar_n}{z^{-n+1}}\right)\ket{\alpha,\alphabar},\nn\\
h_K\cdot\overline\partial\,\phibar \,C_\Lambda\ket{\alpha,\alphabar}& =& -i
C_\Lambda
\left(\frac{h_K\cdot \alphabar}{\zbar} 
+\sum_{n<0}\frac{h_K\cdot \abar_n}{\zbar^{n+1}}+\sum_{n<0}\frac{h_K\cdot\Lambda
 a_n}{\zbar^{-n+1}}\right)\ket{\alpha,\alphabar},
\eea
which implies
\beq \Delta_0^{KK}=(-i)^K P_0\,\left[\C S_K(Q(z))-(-1)^K\C S_K
\left(\Lambda Q\left(\frac{1}{\zbar}\right)\right)\right],\label{Done}\eeq
where
\beq Q(z)=\sum_{n<0}\frac{ a_n}{z^{n}}+\sum_{n<0}\frac{\Lambda
 \abar_n}{z^{-n}}\eeq
and $P_0$ projects out terms $O(z^0)$. Evaluating (\ref{Done}) gives
\bea \Delta_0^{KK}&=&0,\quad\hbox{if $\Lambda=-I$},\nn\\
&=&0,\quad\hbox{if $\Lambda=S$ and $K$ even},\nn\\
&\ne&0,\quad\hbox{if $\Lambda=S$ and $K$ odd}.\eea
 Thus the states $\ket{B(\alpha;S)}$
violate $W_K$
 conservation for all odd $K$. To determine whether this is also the 
case for even $K>2$ we have to  examine the terms
 containing $K-1$ factors $a_n$ with $n\ne 0$.

The operator product $T(w)u_K(z)$ contains a piece proportional to
$(w-z)^{-3}$, and containing $K-1$ factors $a_n, n\ne 0$, which it is known
can be cancelled by forming the combination
\cite{}
\beq u_K'= u_K-2i\alpha_0(\frac{N-K+1}{2})\partial u_{K-1}.\label{ccan}\eeq
Now terms in $W_K$ of the form $b_k \partial W_{K-k-1}W_k,\,k=2\ldots$
 can also 
generate such pieces in the OPE with $T$; but the complete cancellation
(\ref{ccan}) means that these contributions must cancel among themselves.
Therefore, by the linear independence property (\ref{Sindep}), $b_k=0$
and $\Delta_0^{K\,K-1}$ must come from $u_K'$.
Both contributions to $u_K'$ contain $\partial^2\phi$ which
 in terms of the modes satisfies
\bea \partial^2\phi\, C_\Lambda&=&
iC_\Lambda\left(-\frac {a_0}{z^2} -\frac{Q}{z^2}+ \frac{1}{z}Q'(z)\right),\nn\\
\overline\partial^2\phibar\, C_\Lambda&=&
iC_\Lambda\left(-\frac {\abar_0}{\zbar^2} -\frac{\Lambda Q}{\zbar^2}-
\frac{1}{\zbar}\Lambda Q'\left( \frac{1}{\zbar} \right )\right).
\eea
This shows that for each holomorphic sector contribution to $\Delta_0^{K\,K-1}$
 of the form 
$(h\cdot Q )^{K-2} h\cdot Q'$ there is one
 $-(h\cdot\Lambda Q )^{K-2} h\cdot\Lambda Q'$ from
 the anti-holomorphic sector.  Thus when $\Lambda=-I$  they
cancel. On the other hand if $K$ is even and $\Lambda=S$ they add up.
The remaining terms involve just $Q$ and $a_0$ or $\abar_0$
and explicit calculation for this combination  yields
\bea \Delta_0^{K\,K-1}=
 P_0\sum_L&&\left[(h_L\cdot Q(z))^{K-1}
\left(h_L\cdot\alpha-2\alpha_0(\half(N+1)-L)
\right)\right.\nn\\
&&\left.-(-1)^K 
\left(h_L\cdot\Lambda Q\left(\frac{1}{\zbar}\right)\right)^{K-1}
\left(h_L\cdot\alphabar-2\alpha_0(\half(N+1)-L)\right)\right]
\eea
which implies that 
\bea \Delta_0^{K\,K-1}&=&0,\quad\hbox{if $\Lambda=-I$ provided (\ref{alpsym})
},\nn\\
&\ne&0,\quad\hbox{if $\Lambda=S$}.\eea
 This completes the proof that the states
 $\ket{B(\alpha;S)}$
violate $W_K$
 conservation for all $K>2$.
It also shows that the condition (\ref{alpsym}) is necessary,
 although not sufficient, for the $W_K$
to be conserved by the states $\ket{B(\alpha;I)}$.

\section{Discussion\label{DISCUSS}}

To discuss the boundary states of a CFT it is first necessary to know the
primary 
fields. Previous attempts to find the minimal primary field content of
$W_N(p',p)$ 
models using closure of the fusion algebra lead to the prescription
\ref{alphadef} which, 
as we have argued, has some unsatisfactory characteristics. In particular it
does
not distinguish in any reliable way between distinct primary fields and  members
of
the  Felder orbit of a given primary. By requiring modular covariance we have
found the
classification \ref{PFrules}; this leads to an $S$ matrix which satisfies all 
the necessary properties (notwithstanding the proofs given in section
\ref{MODPROPS} we have checked 
this explicitly for examples up to and including rank 4) and the ambiguity in
the specification
of boundary states which was implicit in \cite{Caldeira:2003zz} is removed. A
basis for the boundary states
can be constructed using coherent states which are specified by a primary field
label and a 
generator of the group of outer automorphisms of the Dynkin diagram of $A_r$
(the reader should note
that this was derived, not assumed).  Those corresponding to the identity
conserve the entire
$W$ algebra while those corresponding to the longest element of the Weyl group
maximally break the
chiral algebra down to Virasoro.

For rank 2 the unitary $W_3(p+1,p)$ theories have, in addition to the usual
Cardy
states, an extra $W$ violating physical boundary state for each self-conjugate 
primary field. Each of these states corresponds to an extra physical boundary
condition for
which the annulus amplitude is an admissible partition function with positive
definite Boltzman weights.
There is a straightforward generalization of the Verlinde formula to the
symmetry violating sector. In sum
the results are essentially similar to the Potts case and to those found in some
other examples
of CFTs with extended chiral symmetries \cite{Fuchs:1999xn}.

\acknowledgments
The support of PPARC grant PPA/G/0/2002/00479
 is acknowledged. We acknowledge useful
conversations with S. Kawai. AFC is supported by FCT
(Portugal) through fellowship SFRH/BD/1125/2000. JFW acknowledges the
hospitality of the Niels Bohr Institute and support from MaPhySto, Denmark.

\appendix

\section{$\C P$ for $W_3(p+1,p)$ is $\C C$}
In this appendix $p'$ always means $p+1$.
We will show that 
\beq \C H'=\{\mu\cdot d_1,\; (p+p')\mu\cdot d_1,\; 2pp'-\mu\cdot d_1,\;
2pp'-(p+p')\mu\cdot d_1;\; \mu\in \C C\}\; \hbox{ mod $2pp'$}\eeq
is the same as $\C H$, the set of all even numbers not equal to a multiple of
$p$ or $p'$ and lying between 0 and $2pp'$. The number of elements in $\C H$
is 
\bea \vert\C H\vert&=&(p-1)^2\quad\hbox{$p$ odd},\nn\\
&=& p(p-2)\quad\hbox{$p$ even}.\eea

If $\mu\in\C C$ then $N=\mu\cdot d_1$ is a positive even number such that 
\bea N&\le& p^2-2p-1\nn\\
N&\ne& np+mp',\quad p>n,m\ge 0.\eea
The second condition can be rewritten
\beq N\ne np+m,\quad 0\le m\le n<p\eeq
from which we can immediately see that allowed $N$s are given by
\beq N=np+m,\quad 0\le n<m<p,\;n\le p-3.\label{Nsoln}\eeq
Some further constraints, which depend on whether $p$ is odd or even,
are needed on $m,n$ to ensure that $N$ is even. These only play a role in
calculating $\vert\C C\vert$, the number of allowed $N$s. Using (\ref{Nsoln})
we find that 
\bea \vert\C C\vert&=&\quarter(p-1)^2\quad\hbox{$p$ odd},\nn\\
&=&\quarter p(p-2)\quad\hbox{$p$ even}.\eea
Thus $\vert\C H\vert=\vert\C H'\vert$ and of
 course $\C H'$ does not contain any multiples of $p$ or $p'$. It remains
to show that all the elements of $\C H'$ are distinct.

Consider
\bea \bar N=(p+p')N\;\hbox{mod $2pp'$}&=&p(2m-n)+m\nn\\
&=&\bar n p+\bar m.\label{Nbar}\eea
Note that $0<\bar m <\bar n$ which means that $\bar N$
 can never be one of the $N$s. Now consider
\bea \bar{\bar N}=2pp'-N=p(2(p-m)+2m-n+1)+(p-m).\eea
Comparing this with (\ref{Nbar}) and noting that $2m-n+1>0$ shows that
$\bar{\bar N}$ can never be one of the $\bar N$s. This completes the proof.

\bibliography{W3bib}
\bibliographystyle{JHEP}
\end{document}